\documentclass[preprint,aps]{revtex4}
\usepackage{graphicx}
\begin{document}

\title{Spin Dependent Tunneling in FM$\mid$semiconductor$\mid$FM structures}

\author{S.~Vutukuri, M.~Chshiev and W.~H.~Butler}

\affiliation{MINT Center, University of Alabama, P.~O.~Box 870209, Tuscaloosa, AL 35487}

\begin{abstract}

Here we show that ordinary band structure codes can be used to understand 
the mechanisms of coherent spin-injection at interfaces between ferromagnets 
and semiconductors. This approach allows the screening of different 
material combinations for properties useful for obtaining high tunneling 
magnetoresistance (TMR). We used the Vienna Ab-initio Simulation Code 
(VASP) to calculate the wave function character of each band in periodic 
epitaxial Fe(100)$\mid$GaAs(100) and Fe(100)$\mid$ZnSe(100) structures. 
It is shown that Fe wave functions of different symmetry near Fermi 
energy decay differently in the GaAs and ZnSe.

\end{abstract}

%===================================================================
%\pacs{85.30.Mn, 05.60.Gg, 73.43.Gn, 73.61.-r}
%===================================================================

\date{\today}
\maketitle

%%%%%%%%%%%%%%%%%%%%%%%%%%%%%%%%%%%%%%%%%%%%%%%%%%%%%%%%%%%%%%%%%%%
% INTRODUCTION
%%%%%%%%%%%%%%%%%%%%%%%%%%%%%%%%%%%%%%%%%%%%%%%%%%%%%%%%%%%%%%%%%%%
%\narrowtext

Recently there has been much interest in spin-dependent tunneling between 
ferromagnetic (FM) electrodes separated by insulator~(I) or semiconductor~(S). 
This interest arises both from a desire to better understand spin-dependent 
transport and because of possible technological applications. It has been 
observed experimentally that the tunneling current through a FM$\mid$I$\mid$FM 
sandwich may depend on the relative alignment of the moments of the ferromagnetic 
electrodes on opposite sides of the barrier~\cite{Moodera,Miyazaki,Parkin1,Parkin2,Moodera1,Moodera2}.

Large magnetoresistance was predicted in recent calculations for certain 
epitaxial tunneling systems~\cite{MacLaren,Butler1,Butler2,Mathon,Dederichs}. 
These predictions were based on a spin filtering effect that may arise 
from the symmetry of the wave functions.  At the Fermi energies of bcc Fe, 
bcc Co and CoFeB, there is a difference in the symmetries of wave 
functions between the majority and minority spin channels.  Specifically 
there is a $\Delta_1$ Bloch state for the majority, but not for the 
minority. For some insulating and semiconducting materials, states with 
this $\Delta_1$ symmetry will decay much more slowly than states with 
different symmetries.  Recently these predictions have been largely 
confirmed~\cite{Parkin3,Yuasa,Hayakawa,Djayaprawira}.

In this paper we consider the symmetric structures Fe(100)$\mid$GaAs(100)$\mid$Fe(100) 
and Fe(100)$\mid$ZnSe(100)$\mid$Fe(100). Because the lattice constant of 
bcc Fe is approximately half that of zinc blend GaAs ($2 a_{Fe}/a_{GaAs}=1.014$) 
and ZnSe ($2 a_{Fe}/a_{ZnSe}=1.011$), they fit very well epitaxially. Here 
we report investigations of the potential for spin-dependent transport by
exploring the effect of wave function symmetry on the decay of Bloch
states within the barrier. In systems with two-dimensional periodicity, 
the wave function symmetry is conserved as the electron traverses the interface.  
We observe that wave functions with different symmetries will decay at 
different rates within the barrier.  These symmetries can be determined 
from the angular momentum composition of the Bloch states.

The interfacial structure is critical to understanding tunneling, 
especially, spin-dependent tunneling.  For the case of bcc 
Fe(100)$\mid$MgO(100)$\mid$Fe(100) and similar systems,  it was important 
to find ways of preventing the incorporation of oxygen into the 
interfacial Fe layer~\cite{Butler2,Yuasa}. Here we have studied three 
different epitaxial interfaces in order to search for the most stable 
interface of Fe(100)$\mid$GaAs(100) and Fe(100)$\mid$ZnSe(100)~\cite{Erwin}. 
The structures are presented in Fig.~1 with following details:

Model A: Atomically abrupt interface of bcc Fe and zinc-blende GaAs;

Model B: Partially intermixed i.e., one Fe atom filling the vacancy site in GaAs lattice;

Model C: Fully intermixed i.e., two Fe atoms filling the vacancy sites in GaAs lattice.

\begin{figure}
\begin{center}
\includegraphics[width=7.76cm]{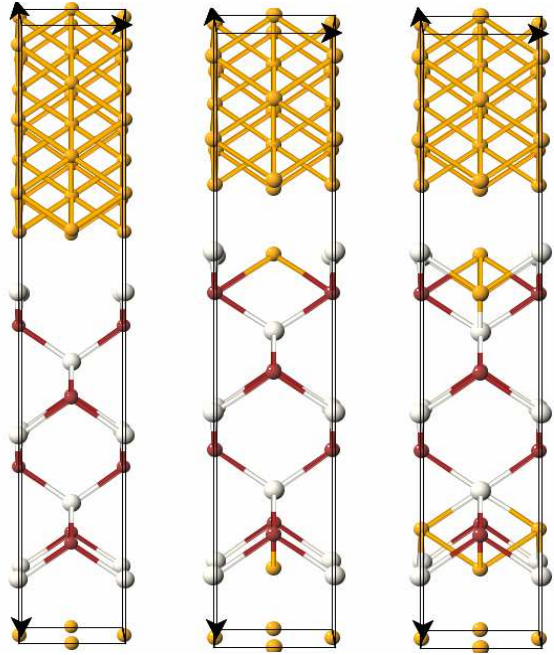}
\caption{Three interface structures of Fe$\mid$GaAs(100) and Fe$\mid$ZnSe(100). 
Models A, B and C (from the left).}
\end{center}
\end{figure}

For each model we attempted first to consider a supercell consisting of 
12 Fe atoms (6 layers for Model A and 5 layers for Models B and C) and 
9 atomic layers of GaAs. It turned out, however, to be impossible to 
construct all models with equal numbers of each type of atom while maintaining 
the same symmetry at both interfaces. To overcome this problem, we constructed 
Models A and C with 14 Fe atoms and 9 atomic layers of GaAs with symmetric 
interfaces. In the case of Model B, the interfacial symmetry requirement 
cannot be fulfilled with 14 Fe atoms. Therefore we approached this problem 
by calculating the energy for this configuration in two different ways. Assuming 
that the effect of the interface will be less in the middle of Fe layer, we have 
calculated the bulk Fe energy taking the interlayer distance at the middle 
of Fe layer. The energy of one layer of Fe from this calculation added to 
the energy of Model B (12 Fe atoms) gives the energy of 14 Fe atoms with 
symmetric interface. In the second case, the energy of Model B with 16 Fe 
atoms was calculated and subtracted from the energy of 12 Fe atoms giving 
thus the energy of 2 Fe layers. Taking half of this energy gives the 
energy of 1 Fe layer (2 Fe atoms per cell per layer), which was added to 
the energy of 12 Fe layers Model B ending up with the energy of 14 Fe 
atoms Model B.  Finally, by comparison of all three models we found that 
the Model A is the most stable, which is consistent with previous 
work~\cite{Erwin}. We performed similar calculations for Fe$\mid$ZnSe structure 
and found again that Model A is more stable than other models.

As a next step, we evaluated the $s$, $p$ and $d$ site projected wave 
function character of bands with different symmetries near the Fermi energy 
for the relaxed structure corresponding to Model A with 14 Fe atoms. 
The calculations were performed using a plane wave based code (VASP)~\cite{Kresse}. 
\begin{figure}
\begin{center}
\includegraphics[width=7.76cm]{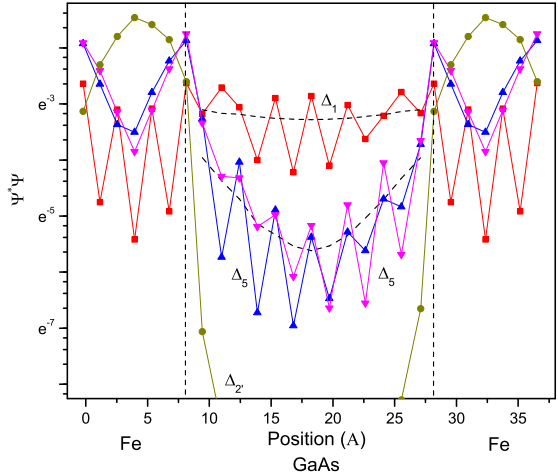}
\caption{Absolute square of $\Delta_1$ (squares), $\Delta_{2^\prime}$ 
(circles) and $\Delta_5$ (triangles) wave functions in a Fe$\mid$GaAs 
supercell. The dashed lines without data points indicate the expected 
decay rate based on Equation~(1).}
\end{center}
\end{figure}
\begin{figure}
\begin{center}
\includegraphics[width=7.76cm]{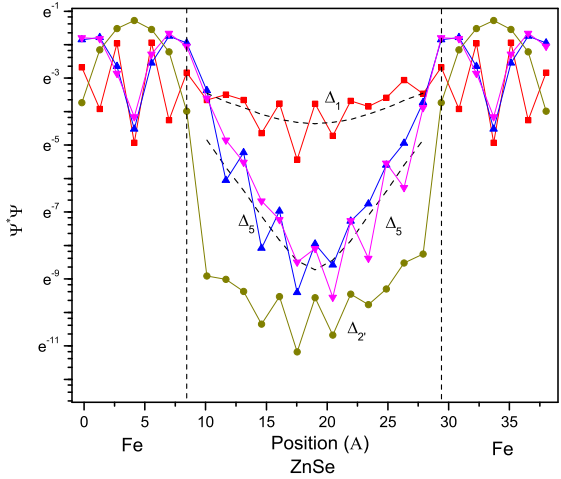}
\caption{The same quantities as in Fig.~2 for Fe$\mid$ZnSe.}
\end{center}
\end{figure}
In Fig.~2 and Fig.~3 we present layer resolved wave function probability 
density, $\psi^*\psi$  for the majority spin state with $\Delta_1$, 
$\Delta_{2^\prime}$ and double degenerate $\Delta_5$ 
symmetry for Fe$\mid$GaAs and 
Fe$\mid$ZnSe structures, respectively. One can see that the slowest 
decay rate is for states with $\Delta_1$ symmetry.  States with $\Delta_5$ 
and $\Delta_{2^\prime}$ symmetry decay much more rapidly. It is clear that 
there is a huge difference in the way wave functions that live primarily 
on the Fe decay into the GaAs and ZnSe.  To clarify the nature of such 
decay rates, we plotted the dependence of the squared quasi momentum $k^2$ 
as a function of energy for Bloch states traveling in the (100) direction 
for GaAs and ZnSe. They are shown in Fig.~4 and Fig.~5, respectively.
In the vicinity of the gap $k^2$ can be represented by
\begin{equation}
\frac{1}{k^2(E)}=\frac{\hbar^2}{2 m_v^* (E-E_v)}+\frac{\hbar^2}{2 m_c^* (E-E_c)}
\end{equation}
where $E_v$ and $E_c$ are the top of the valence band and the bottom of 
the conduction band, respectively, for the $\Delta_1$ band. For both 
of these systems we find that the effective masses $m_v$ and $m_c$ are 
approximately equal at the band edges so that the $k^2$ as a function $E$ 
has the form of a parabola. We have calculated the effective mass $m^*/m$ 
for both GaAs and ZnSe by fitting the above formula to the curves in Fig.~4 and Fig.~5. 
The calculated effective mass, $m^*/m$ is 0.0353 for GaAs and 0.0993 for ZnSe.
The decay of the absolute square of the wave function of a given symmetry 
in the gap will be proportional to $exp(-2|k|z)$,
where $k$ is obtained from 
equation~(1). Note that $k^2$ is negative in the gap so $k$ is imaginary.
\begin{figure}
\begin{center}
\includegraphics[width=7.76cm]{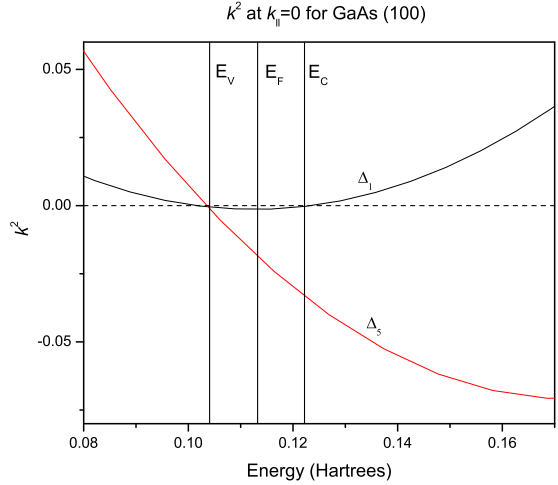}
\caption{Dispersion $k^2(E)$ for GaAs in the vicinity of the gap along $\Delta$~(100). 
$E_v$ labels the top of the valence band and $E_c$ is the bottom of the conduction band.}
\end{center}
\end{figure}
\begin{figure}
\begin{center}
\includegraphics[width=7.76cm]{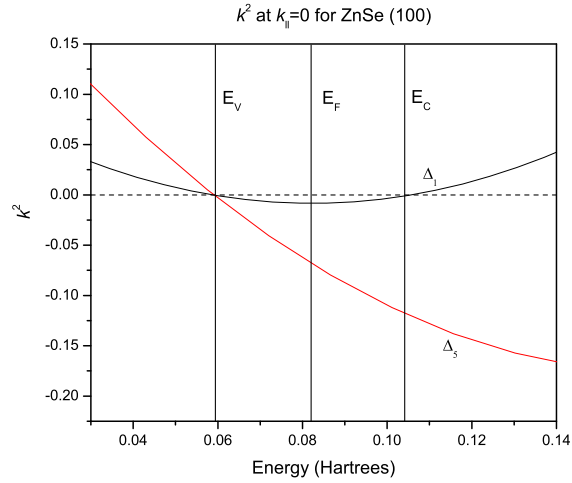}
\caption{The same as Fig.~4 for ZnSe.}
\end{center}
\end{figure} 

In summary, we have shown that Bloch states of only certain symmetries 
are able to propagate through the barrier and the wrong symmetry cannot 
propagate in a metallic electrode. Coherent spin-injection across an 
Fe(100)$\mid$GaAs(100) and Fe(100)$\mid$ZnSe(100) interface can be 
understood using ordinary band structure codes, providing an efficient 
tool to screen material combinations for spin-injection.  It should 
be noted that the energy gaps given by DFT based codes tend to 
significantly underestimate band gaps.  An alternative approach would 
be to use electronic structure calculations to identify the symmetries 
of the complex energy bands at the top and bottom of the gap and then 
to use experimental band masses and energy gap measurements to 
estimate decay rates. 

This work was supported by Information Storage Industry Consortium(INSIC) 
EHDR Program.

\end{document}